\documentclass[prl,aps,twocolumn,showpacs,preprintnumbers,amsmath,amssymb,superscriptaddress]{revtex4}
\usepackage{epsfig,graphics}
\usepackage{dcolumn}
\usepackage{bm}
\usepackage{amsmath}
\usepackage{xcolor}

\newcommand{\brokt}[3]{\left\langle #1 \right| #2 \left| #3\right\rangle} 
\definecolor{dkred}{rgb}{0.6,0.0,0.0}
\definecolor{dkblue}{rgb}{0.123,0.247,0.554}

\begin{document}

\title{Non-magnetic ground state of PuO$_2$}

\author{A. B. Shick}
\affiliation{Institute of Physics, ASCR, Na Slovance 2,
CZ-18221 Prague, Czech Republic}

\author{J. Kolorenc}
\affiliation{Institute of Physics, ASCR, Na Slovance 2,
CZ-18221 Prague, Czech Republic}

\author{L. Havela}
\affiliation{Department of Condensed Matter Physics,  Charles University, Ke Karlovu 5, CZ-12116, Prague, Czech Republic}

\author{T. Gouder}
\affiliation{European Commission, Joint Research Centre, Institute
for Transuranium Elements, Postfach 2340, D-76125 Karlsruhe,
Germany}

\author{R. Caciuffo}
\affiliation{European Commission, Joint Research Centre, Institute
for Transuranium Elements, Postfach 2340, D-76125 Karlsruhe,
Germany}
\date{\today}

\begin{abstract}

The correlated band theory implemented as a combination of the local
density approximation with the exact diagonalization of the Anderson
impurity model is applied to PuO$_2$. We obtain an insulating electronic
structure consistent with the experimental photoemission spectra. The
calculations yield the band gap of 1.8 eV and
a non-magnetic singlet ground state that is characterized by a non-integer
filling of the plutonium $f$ shell ($n_f\approx 4.5$). Due to sizeable
hybridization of the $f$ shell with the $p$ states of oxygen, the
ground state is more complex than the four-electron Russell--Saunders
${}^5{\rm I}_4$ manifold split by the crystal field. The inclusion of
hybridization improves the agreement between the theory and experiment
for the magnetic susceptibility.
\end{abstract}

\maketitle


In order to fully utilize the potential of nuclear power, maintaining at the minimum level the risks associated with the 
deployment of this technology, it is necessary to solve the problems of characterization, treatment, and disposal 
of high-level nuclear waste. On the time scale of several hundred years, the waste
from the open fuel cycle will predominantly contain Pu and minor
actinides. Their geological disposal requires a waste handling
technology of exceptional durability, with highly reduced risk of
accidental events. That is why the comprehensive knowledge of
the physical and chemical properties of actinide-based oxides (AnO$_2$, An = Th, U, Np, Pu, Am, Cm), 
 which 
constitute the main part of the the long-lived nuclear waste, remains a key topic of
condensed matter theory.

PuO$_2$ crystalises in the well-known CaF$_2$ fluorite structure, with eight-coordinated Pu, and four-coordinated O.
For the divalent oxygen, the stoichiometry implies 5$f^4$ configuration for   Pu$^{4+}$. PuO$_2$ is an insulator with 
a band gap of 1.8 eV~\cite{mcneilly1964} and a temperature independent magnetic
susceptibility~\cite{raphael1968}. Recent nuclear magnetic resonance
studies suggest a vanishing local magnetic moment in this compound \cite{yasuoka2012}.

Whilst experimentally the absence of magnetism is clear, its theoretical understanding remains controversial.
The crystal-field (CF) theory \cite{Santini2009} explains this non-magnetic behaviour in terms of  a $\Gamma_1$ nonmagnetic 
singlet ground state, which results from the CF  splitting of  the $J=4$ (${}^5{\rm I}_4$) manifold. The CF picture is consistent
with the inelastic neutron scattering spectra~\cite{kern1999} observing a single peak, corresponding to the $\Gamma_1 \rightarrow \Gamma_4$ transition,  at  the energy of 123 meV. However,  the measured value of the magnetic susceptibility $\chi(T)$  is only 50 \% of 
what one would expect from the Van Vleck coupling,  and its temperature dependence is weaker than the one predicted by the CF model. The average value of $\chi$ in the temperature interval up to about 1000 K can be reproduced, if the 
 $\Gamma_4$ level is taken at 284 meV and not at 123 meV as
 observed. Several alternative mechanisms that could decrease the
 magnitude of the susceptibility while keeping the $\Gamma_1 \to
 \Gamma_4$ gap at 123 meV have been proposed. One of them is an
 effective reduction of the orbital moment by Pu~$f$ -- O~$p$ covalency
 \cite{kern1999}, another involves a negative contribution to $\chi$
 due to antiferromagnetic Weiss exchange field (see
 e.g. \cite{Santini2009} and references therein). Nevertheless, the temperature
 independence of the susceptibility is not explained in these models.

The band-theoretical modeling of the electronic, structural, and
magnetic character  of  actinide materials and their $5f$ states is very
difficult. The conventional density functional theory (DFT) in the
local spin density (LSDA) and generalized gradient (GGA)
approximations falls short to explain the insulating character of
PuO$_2$ as well as other actinide oxides~\cite{wen2013}. It is now
widely accepted that in order to successfully model the actinide
materials, the electron correlations need to be accounted for beyond
the conventional DFT. One of the possibilities is provided by the
so-called hybrid exchange-correlation functionals~\cite{wen2013}.  Unfortunately, these calculations yield the anti-ferromagnetic ground state in disagreement with experiments. 

Contrary to the hybrid functionals, the so-called LSDA+Hubbard $U$ (LDA+U) functional can produce an insulating non-magnetic
solution for PuO$_2$~\cite{nakamura2010, suzuki2013}. However, this
solution is not the minimum energy state, and ferromagnetic and
anti-ferromagnetic spin-polarized LDA+U solutions are lower in
energy. Thus the Pu atom magnetic moment is not quenched, and the experimentally observed temperature independent magnetic susceptibility of PuO$_2$ is not explained by LDA+U calculations. 

In this paper, we extend the LDA+U method making use of a combination
of LDA with the exact diagonalization of the Anderson impurity model
(ED) ~\cite{J.Kolorenc2012, shick13}. We show that the LDA+ED
calculations with the Coulomb $U=6.5$ eV and the exchange $J=0.5$ eV
yield a non-magnetic singlet ground state with $f$-shell occupation
$n_f\approx4.5$ at the Pu atoms. The non-integer filling of the $f$ shell is a
consequence of a hybridization with the $p$ states of oxygen. In
contrast, the ionic bonding with formally divalent oxygen assumed in
the crystal-field theory would require an integer filling ($n_f=4$). The
ground state is found to be separated from the first excited triplet
state by 126 meV. The LDA+ED electronic structure is insulating with a
band gap of 1.8 eV and the calculated density of states (DOS) is
consistent with the experimental results of photoelectron spectroscopy (PES).

The starting point of our approach is the multi-band Hubbard
Hamiltonian $H = H^0 + H^{\rm int} $, where 
$H^0 $ is the one-particle Hamiltonian found from \textit{ab initio} electronic
structure calculations of a periodic crystal; $H^{\rm int}$ is the
on-site Coulomb interaction~\cite{A.I.Lichtenstein1998} describing the
5$f$-electron correlation. We use the LDA for the electron interactions in
other than $f$ shells. The effects of the interaction Hamiltonian
$H^{\rm int}$ on the electronic structure are described with
the aid of an auxiliary  impurity model  describing the complete
seven-orbital 5$f$ shell. This multi-orbital  impurity model  includes the full spherically symmetric
Coulomb interaction, the spin-orbit coupling (SOC), and the crystal
field. The corresponding Hamiltonian can be written as \cite{Hewson}
\begin{align}
\label{eq:hamilt}
H_{\rm imp}  = & \sum_{\substack {k m m' \\ \sigma \sigma'}}
 [\epsilon^{k}]_{m m'}^{\sigma \; \; \sigma'} b^{\dagger}_{km\sigma}b_{km'\sigma'}
 +\sum_{m\sigma} \epsilon_f f^{\dagger}_{m \sigma}f_{m \sigma}
\nonumber \\
& + \sum_{mm'\sigma\sigma'} \bigl[\xi {\bf l}\cdot{\bf s}
  + \Delta_{\rm CF}\bigr]_{m m'}^{\sigma \; \; \sigma'}
  f_{m \sigma}^{\dagger}f_{m' \sigma'}
\nonumber \\
& +  \sum_{\substack {k m m' \\ \sigma \sigma'}}   \Bigl(
[V^{k}]_{m m'}^{\sigma \; \; \sigma'}
 f^{\dagger}_{m\sigma} b_{km' \sigma'} + \text{h.c.}
  \Bigr)
\\
& + \frac{1}{2} \sum_{\substack {m m' m''\\  m''' \sigma \sigma'}}
  U_{m m' m'' m'''} f^{\dagger}_{m\sigma} f^{\dagger}_{m' \sigma'}
  f_{m'''\sigma'} f_{m'' \sigma},
\nonumber
\end{align}
where $f^{\dagger}_{m \sigma}$ creates an electron in the 5$f$ shell
and $b^{\dagger}_{m\sigma}$ creates an electron in the ``bath'' that
consists of those host-band states that hybridize with the impurity
5$f$ shell. The energy position $\epsilon_f$ of the impurity level,
and the bath energies $\epsilon^{k}$ are measured from the chemical
potential $\mu$. The parameter $\xi$ specifies the strength of the
SOC and $\Delta_{\rm CF}$ is the crystal-field potential at the
impurity. The parameter matrices  $V^{k}$ describe the hybridization
between the 5$f$ states and the bath orbitals at energy
$\epsilon^{k}$.

The band Lanczos method~\cite{J.Kolorenc2012} is employed to find
the lowest-lying eigenstates of the many-body Hamiltonian $H_{\rm
imp}$ and to calculate the one-particle Green's function $[G_{\rm
imp}(z)]_{m m'}^{\sigma \; \; \sigma'}$ in the subspace of the $f$
orbitals at low temperature ($k_{\rm B}T=\beta^{-1} =1/500$ eV). 
Then, with the aid of the local Green's
function $G_{\rm imp}(z)$, we evaluate
the occupation matrix
$n_{\gamma_1 \gamma_2} = \frac 1\beta \sum_\omega [G_{\rm imp}(i\omega )]_{\gamma_1 \gamma_2}
\; + \; \frac 12\delta _{\gamma_1 \gamma_2}$, where the composite
index $\gamma\equiv (l m \sigma)$ labels spinorbitals.
  
The matrix $n_{\gamma_1 \gamma_2}$ is used to construct an effective LDA+$U$
potential ${V}_{U}$, which is inserted into Kohn--Sham-like
equations~\cite{shick09}:
\begin{gather}
\bigl[ -\nabla^{2} + V_{\rm LDA}(\mathbf{r}) + V_{U} + \xi ({\bf l} \cdot
{\bf s}) \bigr]  \Phi_{\bf k}^b({\bf r}) = \epsilon_{\bf k}^b \Phi_{\bf
k}^b({\bf r}).
\label{eq:kohn_sham}
\end{gather}
These equations are iteratively solved until self-consistency over
the charge density is reached. In each iteration a
new value of the 5$f$-shell occupation is obtained from the
solution of Eq.~(\ref{eq:kohn_sham}), and the next iteration is started by
solving~Eq.~(\ref{eq:hamilt}) for the updated 5$f$-shell
filling. The self-consistent procedure  was repeated until the
convergence of  the 5$f$-manifold occupation $n_f$ was better than
0.02.

Once the self-consistency is reached, the eigenvalues $ \epsilon_{\bf k}$
of Eq.~(\ref{eq:kohn_sham}) are corrected to account for the selfenergy $\Sigma(\epsilon)$
of the impurity model Eq.~(\ref{eq:hamilt}), see supplementary material for 
additional details.  We make use of the first-order perturbation theory
 to write an eigenvalue correction,
\begin{equation}
\label{eq:qp}
 {\rm E}_{\bf k}^{n}=\epsilon_{\bf k}^n + \mathop{\rm Re}\brokt{\Phi^n_{\bf k}}{\Sigma({\epsilon_{\bf k}^n})-V_{U}} {\Phi_{\bf k}^n}.
\end{equation}

The SOC parameters $\xi=0.30$ eV for
PuO$_2$  was determined from LDA calculations. 
CF effects were described by the crystal field potential for
the cubically coordinated $f$-shell,
\begin{eqnarray}
\Delta_{\rm CF}&=& \frac{16\sqrt{\pi}}{3}  V_{4} \biggl(Y_{4}^{0} +
\sqrt{\frac{10}{7}} \mathop{\rm Re} Y_{4}^{4}\biggr) \nonumber \\
 &+&   32 \sqrt{\frac{\pi}{13}} V_{6} \bigl(Y_{6}^{0} - \sqrt{14}
 \mathop{\rm Re} Y_{6}^{4}\bigr)\,, \label{eq:CF} 
\label{eq:cf}
\end{eqnarray}
where $V_{4}$ and $V_{6}$ were chosen as external parameters. 
In the actual calculations, we used the values $V_4=-0.151$ eV and $V_6=0.031$ eV deduced from experimental data in Ref.~\cite{kern1999}, and close to the estimate given in \cite{Magnani2005}. The CF parameters could be also calculated using ab-initio approaches~\cite{pnovak2013},
and we plan to do so in the future.

\begin{figure} 
\includegraphics[angle=0,width=0.975\columnwidth]{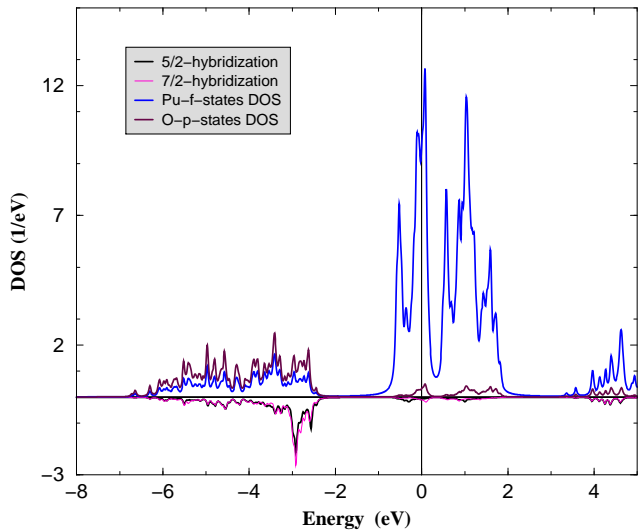}
\caption{(Color online) The  O-$p$ and Pu-$f$-projected
DOS, and the hybridization
function ${\Delta}(\epsilon)$ (the negative $y$-axis scale, eV). } \label{hybridization}
\end{figure}

 In order to specify the bath parameters, we
assume that LDA represents the non-interacting model for PuO$_2$,
and associate with it the solution of Eq.~(\ref{eq:hamilt}) without
the last Coulomb-interaction term. Moreover, we assume that
the first and fourth terms in
Eq.~(\ref{eq:hamilt}) are diagonal in $\{j,j_z\}$ representation.
Next, we obtain $V_{k=1}^{j=5/2,7/2}$ and $\epsilon_{k=1}^{5/2,7/2}$
from LDA
hybridization function ${\Delta}(\epsilon) =  -{1 \over {\pi N_f}} \mathop{\rm
  Im}\mathop{\rm Tr}
[G^{-1}(\epsilon + i \delta)]$ where $N_f=6$ for $j=5/2$,
$N_f=8$ for $j=7/2$, and $G$ is the LDA Green's function.
The hybridization function ${\Delta}(\epsilon)$ is shown in Fig.~\ref{hybridization} together
with the O-$p$ and Pu-$f$-projected LDA densities of states.
As follows from Fig.~\ref{hybridization}, the most essential hybridization
occurs in the energy region of the O-$p$ states. We set  $\epsilon_{k=1}^{5/2,7/2}$ to
the $-2.92$ eV  peak position of  ${\Delta}(\epsilon)$, 
and obtain $V_{k=1}^{j=5/2}=1.46$  eV,  and $V_{k=1}^{j=7/2}=1.62$  eV 
at the peak position of  ${\Delta}(\epsilon)$.

The Slater integral $F_0$ (Coulomb $U$) is regarded as an adjustable parameter; 
calculations have been performed for $U= 4.5$, 5.5, and 6.5 eV, within
the range commonly considered in the literature. For the other Slater
integrals we have used the values $F_2=5.96$ eV, $F_4=3.982$ eV, and
$F_6= 2.946$ eV that have been obtained by scaling the atomic Hartree-Fock
results~\cite{KMoore2009} to approximately 60\% to account for
configuration interactions and screening effects. The
screened integrals correspond to the Hund's exchange $J = 0.5$ eV, which is in
the ballpark of  the values  used in the LDA+U~\cite{wen2013} and 
LDA+DMFT~\cite{Yin2011} calculations. 
For the  double-counting term (included in  the potential $V_U$)  we have adopted the fully-localized (or
atomic-like) limit (FLL)  $V_{dc} = U (n_f-1/2) - J(n_f-1)/2$.

In the calculations we used  an in-house
implementation~\cite{shick99,shick01} of the full-potential linearized augmented plane wave (FP-LAPW) method
that includes both scalar-relativistic and spin-orbit coupling effects. The calculations were
carried out assuming a paramagnetic state, and the cubic fluorite crystal structure. We set the radius of the Pu atomic sphere
to 2.65~a.u, and the O atomic sphere to 1.70 a.u. The parameter
$R_{\rm Pu} \times K_{\text{max}}=9.3$ determined the
basis set size, and the Brillouin zone was sampled
with 4000 $k$~points. 

Now  we turn to the results of  LDA+ED calculations. 
For the  set of $U=4.5$ eV and $J=0.5$ eV, solving self-consistently Eq.~(\ref{eq:hamilt}) and Eq.~(\ref{eq:kohn_sham}),
we obtain the $f$ occupation $n_f=4.58$ close to conventional LDA+U
with the same $U$ and $J$ ($n_f=4.56$)  as well as to the occupation deduced from
the $4f$ X-ray photoemission spectroscopy (XPS, $n_f=4.65$ \cite{kotani1992}). After applying the eigenvalue correction 
Eq.~(\ref{eq:qp}), we do not obtain an insulating  state. Once the
Coulomb $U$ is increased, say to  5.5~eV, the  PuO$_2$ becomes an insulator
with the band gap of  1.4~eV (see supplementary material, Table S2). For the Coulomb $U=6.5$~eV (and $J=0.5$~eV),  we obtain an insulating solution
with a  band gap of 1.8~eV. When the value of $J=0.6$~eV is used, the  band gap value is slightly reduced to  1.6~eV. 
The corresponding total density of states (TDOS),  the Pu atom $f$-state, and the O atom $p$-state partial DOS are shown 
in Fig.~\ref{fig:dos2}.

 \begin{figure} 
\centerline{ \includegraphics[angle=0,width=0.975\columnwidth]{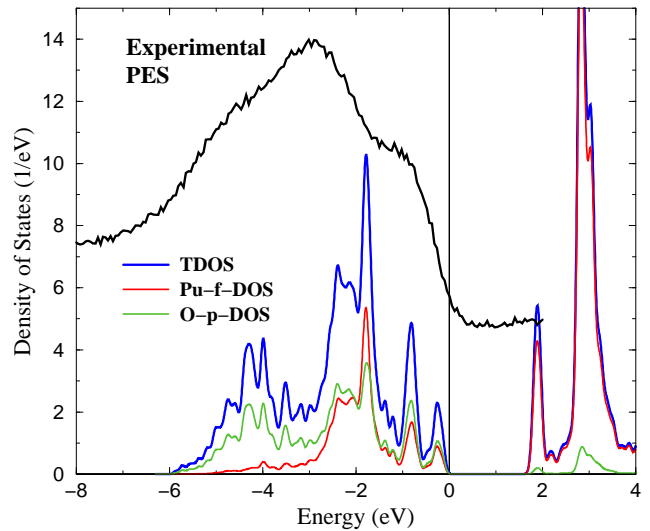}}
\caption{(Color online) 
The total, O-$p$ and Pu-$f$-projected
DOS from LDA+ED calculations with $U=6.5$ eV, $J= 0.5$ eV, 
together with the experimental PES (spectrum, recorded with the He-II excitation, photon energy 40.81 eV). Note that the PES spectrum 
is adjusted to match the upper edge with the zero energy.}
\label{fig:dos2}
\end{figure}

The experimental PES~\cite{Butterfield2004, Gouder2007, Gouder2013}
(see Fig.~\ref{fig:dos2}, note the horizontal shift of the data)
is usually obtained on PuO$_2$ films prepared by reactive sputter deposition from an $\alpha$-Pu target in an Ar/O$_2$ plasma. 
The O$_2$ partial pressure was adjusted to obtain the correct stoichiometry.
Peaks are observed at 2 and 4 eV binding energy (BE), with a shoulder at   6 eV BE.  The orbital parentage of the peaks can be 
deduced by comparing the intensities obtained for He-I and He-II
radiation  (photon energy equal to 21.22 eV and 40.81 eV, respectively). 
The He-I spectrum is dominated by the O-$p$ emission, whereas for He-II the Pu-$f$ and O-$p$ contributions are comparable. 
It is concluded that the 2 eV BE peak stems from the  Pu-$f$ states.  This peak is usually considered as an indication of the $f^4$ nominal configuration, corresponding to Pu${}^{4+}$ oxidation state. The next  (4 eV BE) peak is more intense and broad. The He-II and He-I spectral difference indicates a substantial
 O-$p$ character of this peak. The  shoulder at   6 eV BE is associated
 mostly with the O-$p$ states.  As the calculations associate the upper edge of the conduction band with zero binding energy, the experimental spectrum was shifted for the sake of comparison towards the zero energy, as well.

The LDA+ED DOS shown in Fig.~\ref{fig:dos2} has the  peak with the
mixed Pu-$f$ and O-$p$ characters at $\approx$ 0.8 eV below the $E_F$
(with additional smaller satellite closer to the  $E_F$). Another
broad peak at $\approx -2$ eV has more intensity
(for both $f$ and $p$ states). And there is a broad, dominantly  O-$p$ character, shoulder between $-3$ and $-6$ eV. Thus, if we consider the 
difference in the peaks positions, they correspond  reasonably to the experiment. Their absolute values differ from the experimental
BE~\cite{Butterfield2004, Gouder2007, Gouder2013}  by $\approx$ 1 to 1.5 eV. The reason is that in PES experiments the Fermi level falls in the middle of the band gap, whilst the upper edge of the conduction band defines the Fermi level in the calculations.
 Both the experiment and
calculations suggest   a mixture of   $f^4$  and  $f^5$ configurations in the ground state due to Pu-$5f$ and O-$2p$ hybridization.

Now we turn to the salient theme of our investigation, the ground
state of  the  impurity model  Eq.~(\ref{eq:hamilt}). This ground state
is a singlet formed by the 5$f$~shell and the bath with occupation
numbers $\langle n_f \rangle=4.52$ in the $f$~shell and $\langle
n_{bath} \rangle=13.48$ in the bath states.  Since the ground state 
is a singlet, any magnetic or
multipolar degree of freedom is frozen when the temperature is
well below the gap between the ground state and excited states.
The calculated energy difference between
this ground state and the first excited states (triplet) is 126 meV,
very close to the experimental value 123 meV, observed
in the inelastic neutron scattering spectra~\cite{kern1999}. Neither
the ground state nor the excited states are exact crystal-field
states since they involve the $p$ states of oxygen.

Analogously to the crystal-field theory, the impurity model can be
used to estimate the magnetic susceptibility and its temperature
dependence by adding the action of an external magnetic field $B_z$ to
Eq.~(\ref{eq:hamilt}),
\begin{multline}
\label{eq:HB}
\hat H_B =
 - \sum_{m\sigma}\mu_{\rm B} B_z\Bigl(
\bigl[\hat l_z+2\hat s_z\bigr]_{mm}^{\sigma\;\;\sigma}
\hat f_{m\sigma}^{\dagger}\hat f_{m\sigma}\\
+ \bigl[2\hat s_z\bigr]_{mm}^{\sigma\;\;\sigma}\,
\hat b_{m\sigma}^{\dagger}\hat b_{m\sigma}\Bigr)\,.
\end{multline}
The bath originates from the LDA oxygen bands and hence the magnetic
field couples only to the spin degrees of freedom in this part of the
impurity model. When the magnetic field is weak and the
linear-response regime applies we can get the molar magnetic
susceptibility from the induced $f$-shell magnetic moment
$m_z(T)=\mu_{\rm B}\langle \hat l_z+2\hat s_z\rangle$ as
$\chi_{\rm imp}(T)=\mu_0 N_{\rm A} m_z(T)/B_z$, where $\mu_0$ stands
for the vacuum permeability.

\begin{figure}[t]
\centerline{\includegraphics[angle=0,width=0.975\columnwidth]{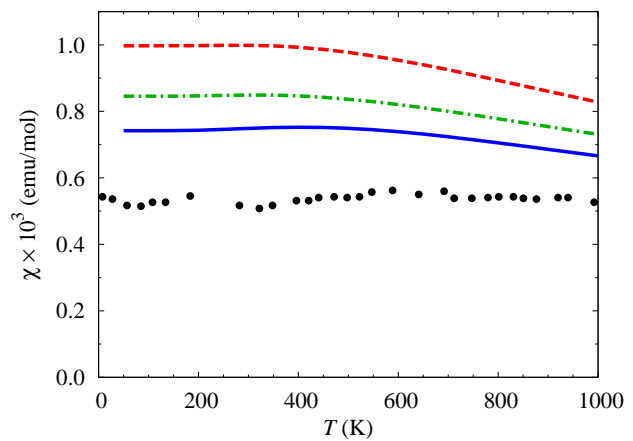}}
\caption{(Color online) 
 Temperature dependence of the molar magnetic susceptibility
 calculated in the crystal-field theory ($\chi_{\rm CF}$, red dashed
 line) and in the impurity model ($\chi_{\rm
   imp}$, green dot-dashed line) for $U=6.5$ eV and $J=0.5$ eV. Blue
 solid line shows $\chi_{\rm imp}$ with the exchange enlarged to
 $J=0.6$ eV. Black dots are the experimental data from
 \cite{raphael1968}.
 }
\label{fig:chi}
\end{figure}

The susceptibility $\chi_{\rm imp}$ calculated for $U=6.5$ eV and
$J=0.5$ eV is shown in Fig.~\ref{fig:chi} in comparison with the
susceptibility $\chi_{\rm CF}$ from the crystal-field theory,
that is, from Eq.~\eqref{eq:hamilt} without the bath. The
hybridization of the $f$ orbitals with the ligand states reduces the
magnitude of the susceptibility as well as its temperature
dependence and hence brings the theory closer to the experimental
findings \cite{raphael1968}. Here we in fact estimate the
magnitude of the covalency effects discussed in
\cite{kern1999}. Furthermore, it turns out that $\chi_{\rm imp}$ is
sensitive to the exchange parameter $J$, whereas it is essentially
independent on $U$. With $J=0.6$ eV, the temperature
dependence is further suppressed and the magnitude of
$\chi_{\rm imp}$ is only about 35\% larger than the experiment
(see Fig.~\ref{fig:chi} and compare to nearly 100\% overestimation of
the crystal field theory alone). The $\Gamma_1\to\Gamma_4$ gap
practically does not depend on~$J$. The reduced temperature dependence is
a result of cancellation of temperature-dependent parts of the induced
moments $ \mu_{\rm B}\langle\hat l_z\rangle$ and
$\mu_{\rm B}\langle 2\hat s_z\rangle$ that both deviate from a constant
above 300 K due to increasing population of the excited-state
triplet $\Gamma_4$. See supplementary material for an
illustration and for additional details of the $J$-dependence of
$\chi_{\rm imp}$.

In summary, making use of the LDA+ED calculations with $U=6.5$ eV and
$J=0.5$ eV we obtain a non-magnetic singlet ground state with
$n_f\approx 4.5$ for Pu atoms in PuO$_2$. The LDA+ED yields an insulating
electronic structure consistent with the experimental photoelectron spectra. 
The band gap is found to be 1.8 eV. The energy difference between
the ground state and the first excited triplet state is 126 meV, in
agreement with the experimental inelastic neutron scattering
spectra. The calculated singlet ground state and the consequent
non-magnetic behavior have a lot in common with the outcome of the
crystal-field theory, the significant improvement is that LDA+ED
achieves these features for a realistic non-integer occupation of the
Pu $f$ orbitals. We emphasize that we did not adjust the model
parameters to fit the experimental
findings. Instead, we investigated the dependence of physically
observable quantities on the choice of these parameters.

\begin{acknowledgements}
The financial support from the Czech Republic Grant GACR
P204/10/0330 is acknowledged.
\end{acknowledgements}

\vspace*{1cm}

\centerline{\textbf{ \em{Supplemental Material for}}}
 
\centerline{\textbf{Non-magnetic ground state of PuO$_2$}}

\section{Analysis of the LDA+U results}

There are a number of recent papers reporting the LDA/GGA+U calculations
for PuO$_2$ (see for instance Ref.~\cite{Jomard2008} and references
therein). They obtain an anti-ferromagnetic ground state, but often
neglect spin-orbit coupling and use oversimplified treatment of the
electron-electron Coulomb and exchange interactions. Nakamura {\em et
al.}~\cite{nakamura2010}, and later Suzuki, Magnani and
Oppeneer~\cite{suzuki2013}, have shown that constrained non-magnetic
LDA+U calculations yield an insulating electronic structure for
PuO$_2$. However, this non-magnetic solution is not the variationally
determined minimum-energy state~\cite{nakamura2010,wen2013}. The
ferromagnetic and antiferromagnetic spin-polarized solutions are lower
in energy and hence the non-magnetic insulating solution of
Refs.~\cite{nakamura2010,suzuki2013} is not the LDA+U ground state.
 
Below we perform an additional analysis of our own calculations using
the conventional LDA+U. For the values of $U=4$ eV and $J=0$ eV our
results are in agreement with those of Suzuki {\em at
  al.}~\cite{suzuki2013}, and show an insulating state with the band gap
of 1.8 eV and the $f$-states occupation $n_f=4.54$. The total density
of states (TDOS), the O atom $p$-state and the Pu atom $f$-state
partial DOS are shown in Fig.~\ref{fig:dos}.

When a realistic non-zero $J=0.5$ eV is included, and $U=4.5$
eV is adopted (this choice keeps the effective $(U-J)= 4.0$ eV), the
unoccupied part of the DOS changes substantially. The narrow empty
$f$-peak shifts down in the energy, and the band gap is reduced to the
value of 1.1 eV. For the realistic value of exchange $J$ (Hund's
exchange), the Coulomb $U=6.5$ eV is needed to recover the band gap of
1.8 eV (see Tab.~\ref{tab:gap} and Fig.~\ref{fig:dos}). Indeed, this
non-magnetic LDA+U solution is not the true ground state for PuO$_2$,
and allowing for the spin polarization will lower the total
energy~\cite{nakamura2010}.

\renewcommand{\thetable}{S1}
\begin{table}[htbp] 
\caption{Occupation $n_f$ of the $f$ shell and the band gap (eV) calculated
  in LDA+U as functions of the Coulomb $U$ and the exchange $J$.}
\begin{ruledtabular}
\begin{tabular}{ccccc}
\mbox{[{$U$, $J$}] (eV)}&\mbox{[4.0, 0.0]}&\mbox{[4.5, 0.5]}&\mbox{ [5.5, 0.5]}&\mbox{ [6.5, 0.5]} \\ 
\hline
$n_f$                   &4.54&4.56&4.52&4.48\\
\mbox{$E_g$ (eV)} &1.8  &1.1   &1.3  &1.8  \\
	\end{tabular}
	\end{ruledtabular}
\label{tab:gap}
\end{table}

\renewcommand{\thefigure}{S1}
\begin{figure*}[htbp]
\centerline{\includegraphics[angle=0,width=0.975\columnwidth]{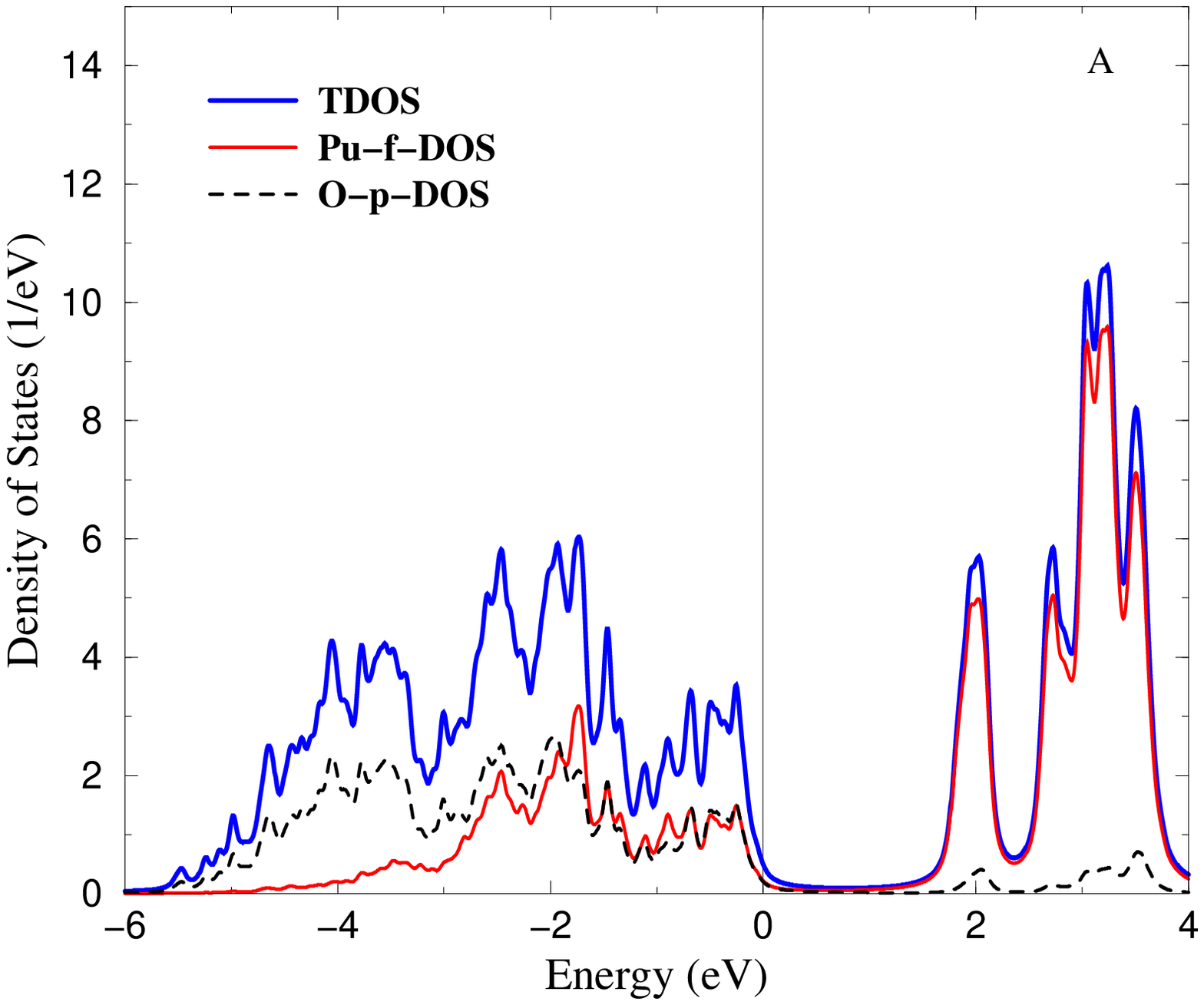} \includegraphics[angle=0,width=0.975\columnwidth]{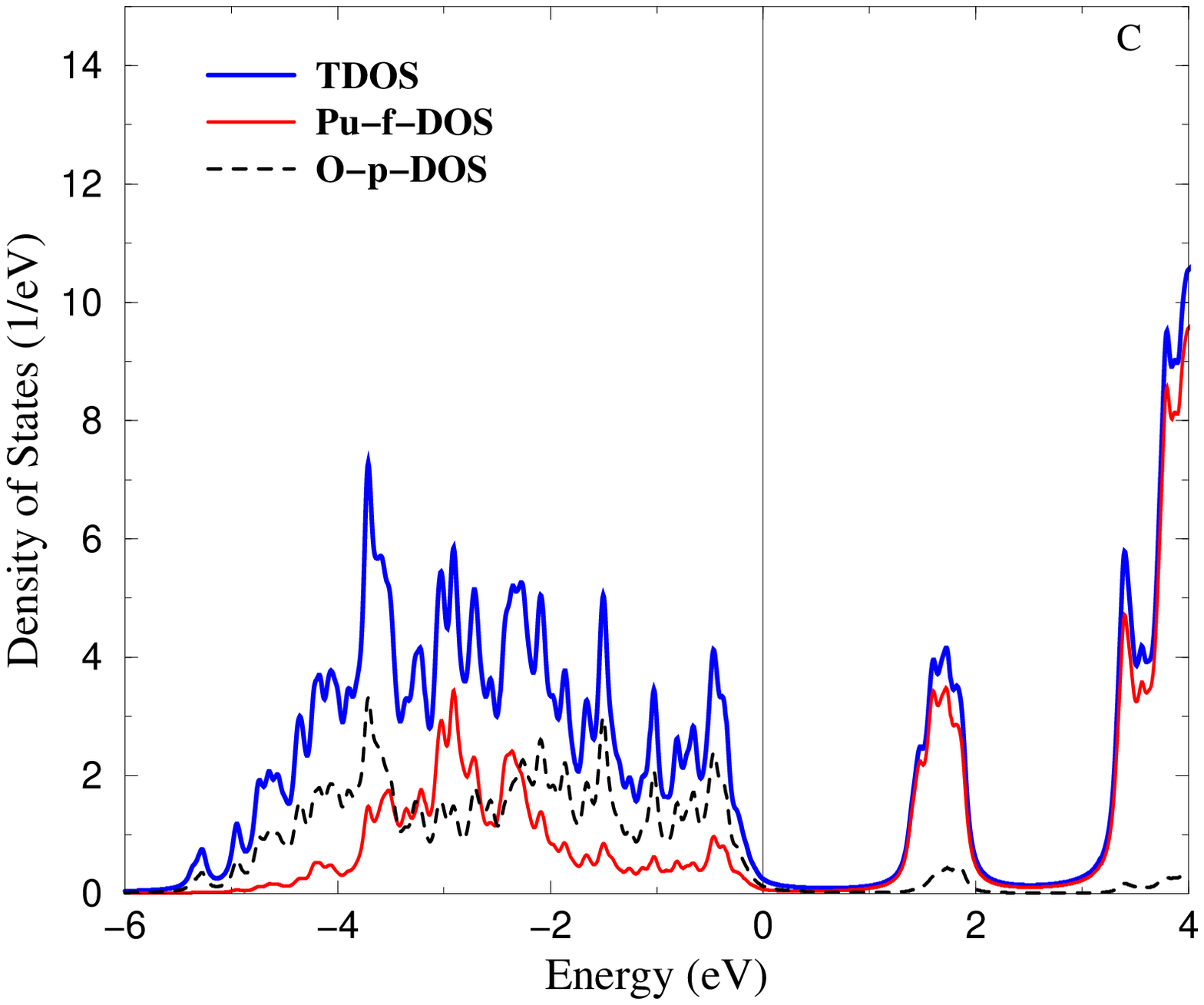}}
\centerline{\includegraphics[angle=0,width=0.975\columnwidth]{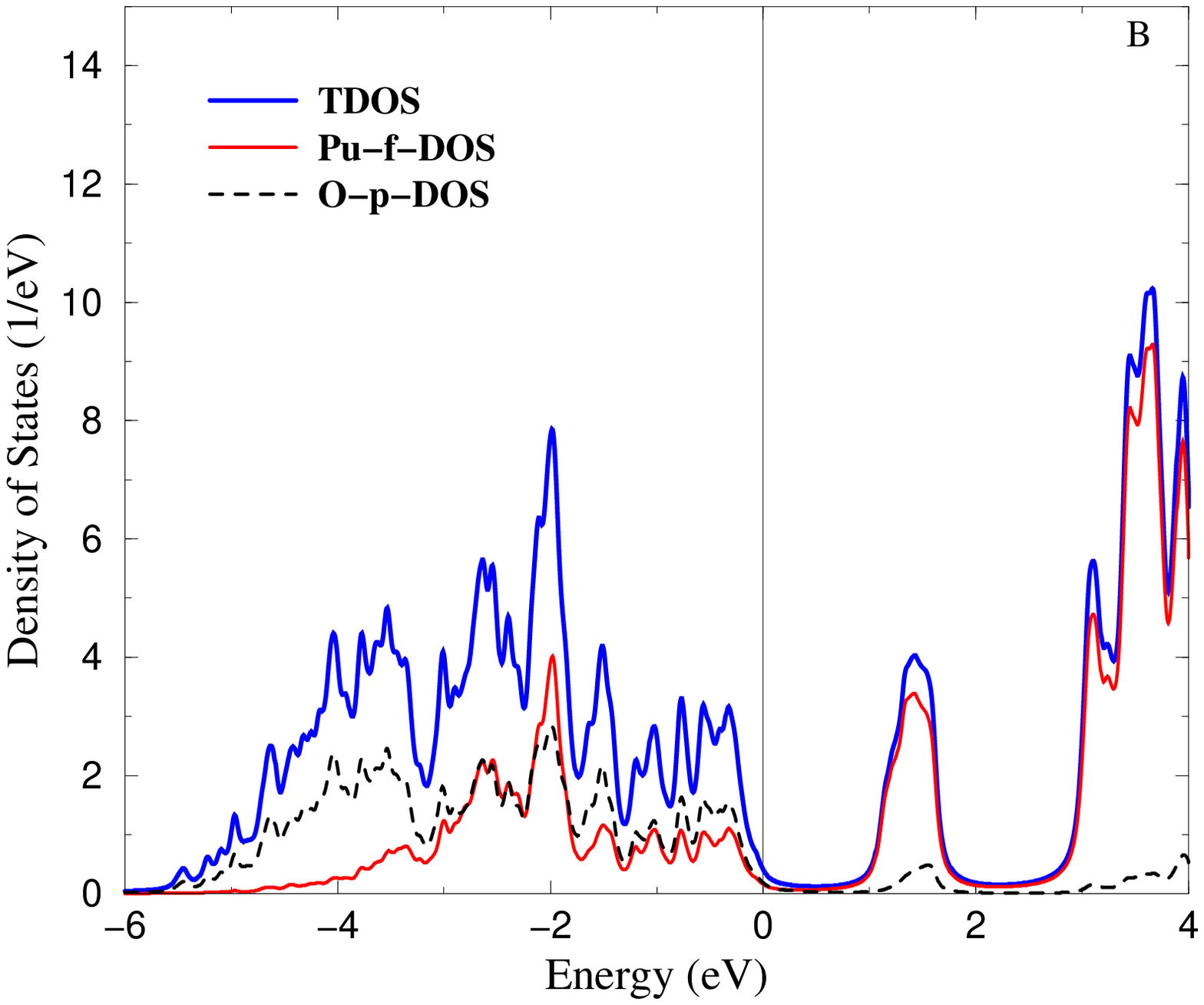} \includegraphics[angle=0,width=0.975\columnwidth]{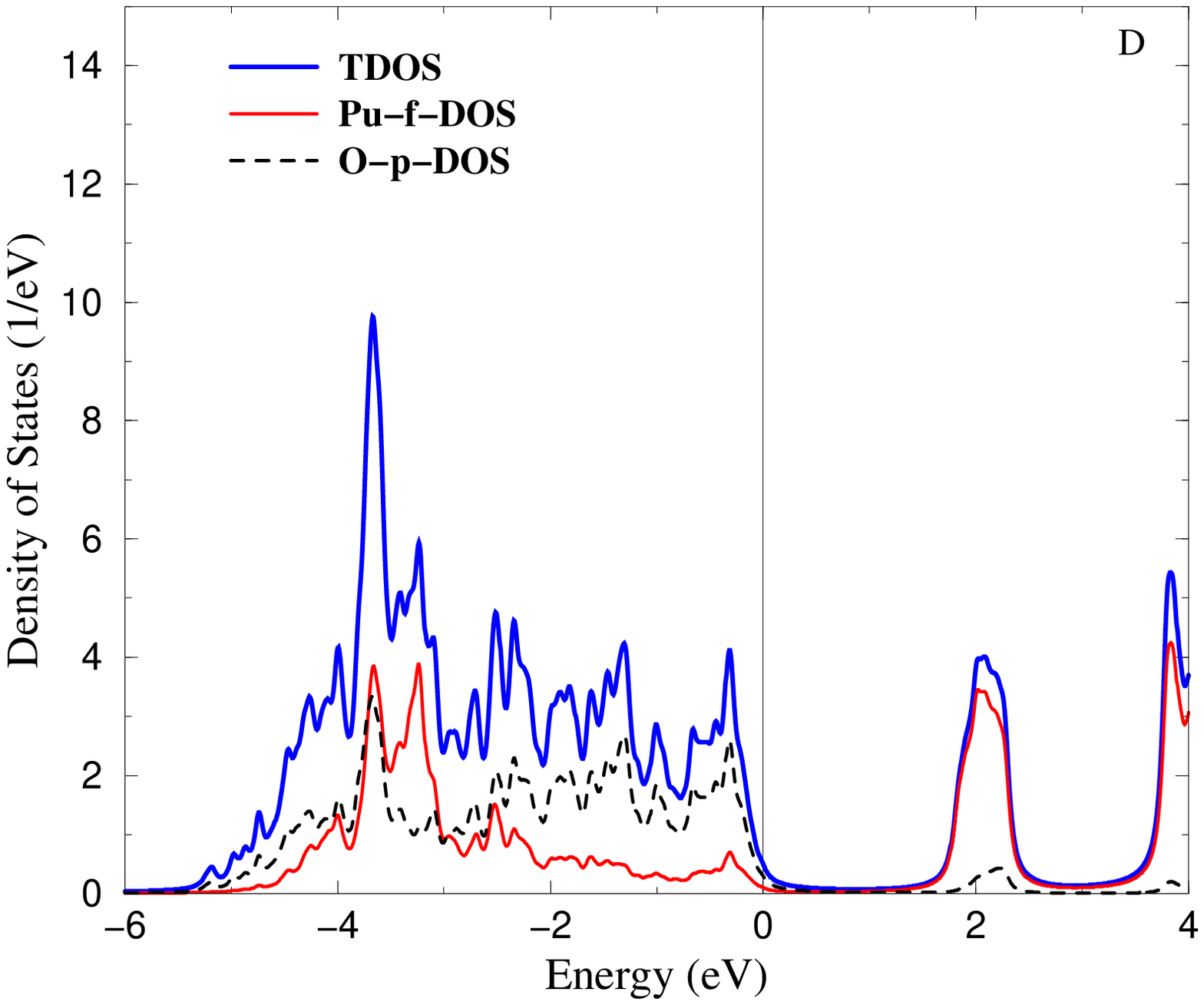}}
\caption{(Color online) 
The total, O-$p$ and Pu-$f$-projected
DOS from LDA+U calculations with $U=4$ eV, $J= 0$ eV (A), and 
 $U=4.5$ (B), 5.5 (C), 6.5 (D) eV, and $J= 0.5$ eV .}
\label{fig:dos}
\end{figure*}

\section{The Anderson impurity model}
The Anderson impurity model including the spin-orbital coupling is
block-diagonal in the occupation-number basis, with each block
corresponding to a particular total number of electrons in the model
$N_{\rm tot}$. The blocks relevant for our problem correspond to
$N_{\rm tot}$ equal 17, 18 (the ground state is in this block) and
19. The model is solved for these fillings with the aid of the Lanczos
method using {\scshape arpack} \cite{arpack} and an in-house
code. Although this technique is efficient, the calculations still
take a lot of time and require a large amount of RAM due to large
dimensions of the blocks, namely 21.4, 13.1 and 6.9 million. In order
to make the problem more manageable, we reduce the many-body basis
using the method of Gunnarsson and Sch\"onhammer \cite{gunnarsson1983}
that is related to the expansion in the hopping integrals $V_k$ around
the atomic limit. For instance, the reduced Fock space containing all
contributions up to the second order of this expansion reads as
$\{|f^n\rangle, |f^{n+1}\underbar{$b$}\rangle,
|f^{n+1}\underbar{$b$}\underbar{$b$}\rangle\}$ where $\underbar{$b$}$
indicates a hole in the bath orbitals. The size of the basis can thus
be specified by the maximum number of bath holes present, $N_{\rm h}$.

Since the hybridization parameters $V_k\approx 1.5$ eV are not
particularly small, it is not obvious if the expansion converges
quickly enough to be actually useful. Explicit convergence analysis
nevertheless shows that $N_{\rm h}=3$ gives essentially converged
$f$-electron spectrum and magnetic susceptibility, whereas
$N_{\rm h}=2$ is still too small. The spectral 
density calculated as $N(\epsilon) = -\mathop{\rm Im}
\mathop{\rm Tr} [G_{\rm imp}(\epsilon + i \delta)]/\pi$ is shown in
Fig.~\ref{fig:dos2}A. There is an approximately 1 eV shift of the spectrum
to the right and the gap is about 0.5 eV smaller when the basis is
increased from $N_{\rm h}=2$ to $N_{\rm h}=3$. No substantial changes
in the spectrum are found with the further increase of $N_{\rm h}$.

The spectral density for $N_{\rm h}=2$ closely resembles the $f$-DOS from Fig.~3 of
Ref.~\cite{Yin2011} that employs the so-called one-crossing
approximation which is also a variant of the expansion around
the atomic limit. The total density of states (TDOS), the Pu atom
$f$-state, and the O atom $p$-state partial DOS are shown in
Fig.~\ref{fig:dos2}B for $N_{\rm h}=2$. There are some differences
with the DOS for $N_{\rm h}=3$ shown in Fig.~2 of the main text. The
band gap becomes slightly larger, $\approx$ 2 eV, for $N_{\rm h}=2$.
The occupied part of the Pu-$f$-DOS is shifted to the left by about
0.5 eV, and the O-$p$-states are following this shift. This is due to
the shift of the SDOS shown in Fig.~\ref{fig:dos2}A. There are no
noticeable changes at the bottom of the valence band where the
O-$p$-states dominate.

  
\renewcommand{\thefigure}{S2}
\begin{figure*}[htbp]
{\includegraphics[angle=0,width=0.975\columnwidth]{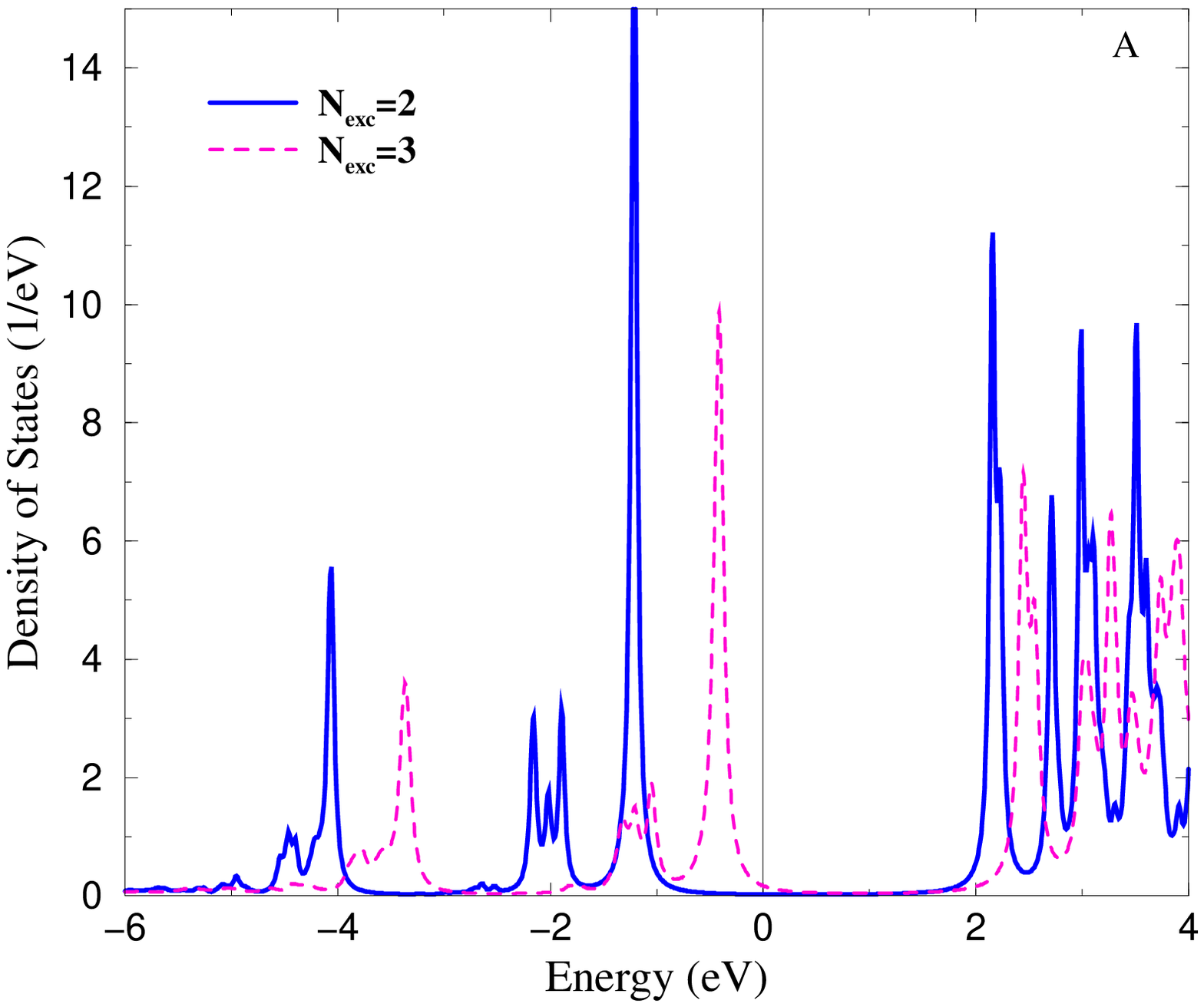}   \includegraphics[angle=0,width=0.975\columnwidth]{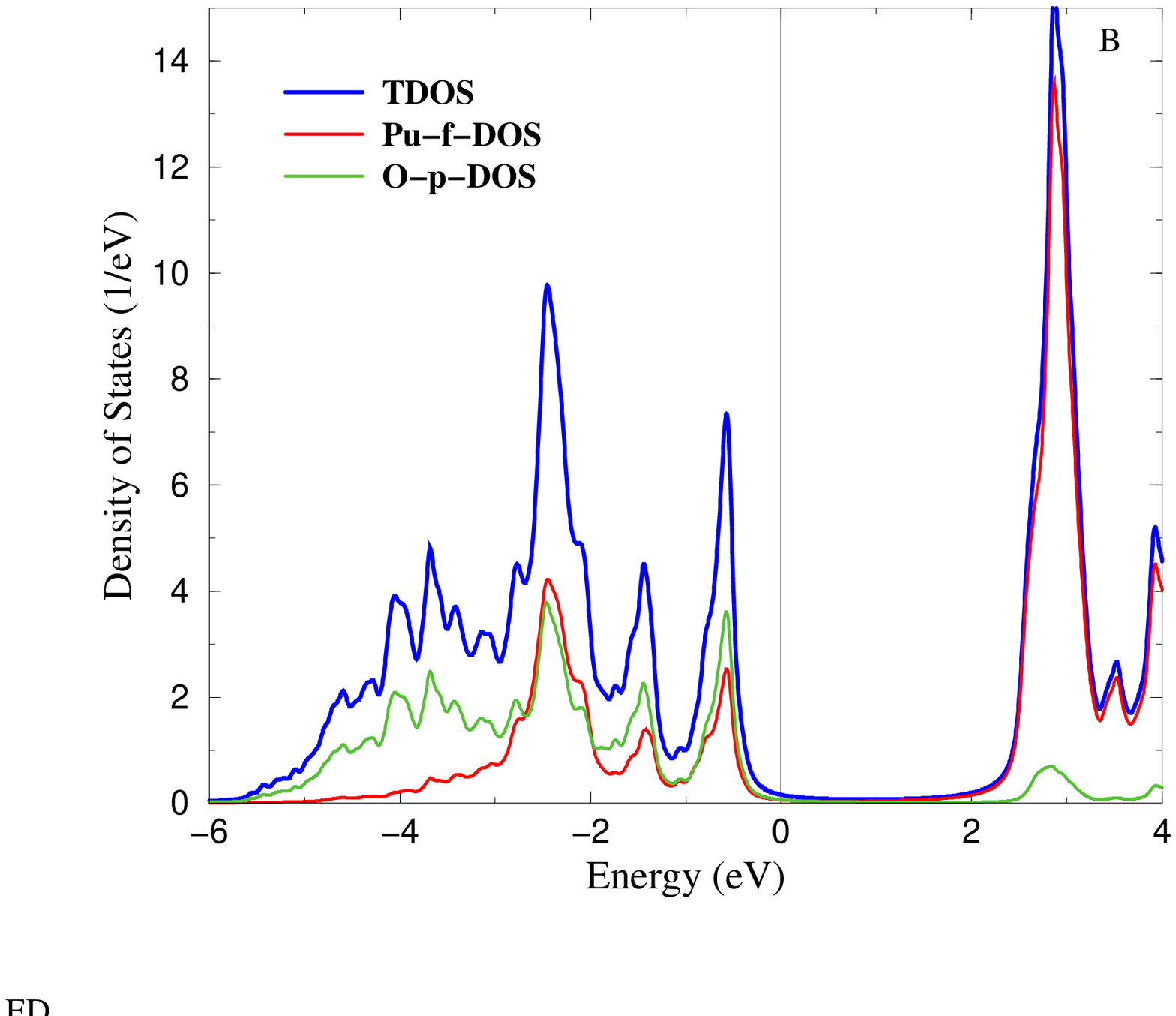}}
\caption{(Color online) 
The spectral $f$-DOS calculated in the impurity model for 
 $U=6.5$ eV, $J= 0.5$ eV evaluated for different sizes of the
 many-body basis (A). The total, O-$p$ and Pu-$f$-projected
DOS from LDA+ED calculations with $N_{\rm h}=2$ (B).}
\label{fig:dos2}
\end{figure*}

\renewcommand{\thetable}{S2}
\begin{table}[htbp] 
\caption{Occupation $n_f$ of the $f$ shell, and the band gap (eV) calculated
in LDA+ED as a function of
the Coulomb $U$, and the exchange $J$.}
\begin{ruledtabular}
\begin{tabular}{ccccc}
\mbox{[{\em U, J}] (eV)}&\mbox{[4.5, 0.5]}&\mbox{ [5.5, 0.5]}&\mbox{ [6.5, 0.5]} \\ 
$n_f$                   &4.58&4.55&4.52\\
\mbox{$E_g$ (eV)} &-  &1.4  &1.8 \\
	\end{tabular}
	\end{ruledtabular}
\label{tab:gapLDAED}
\end{table}

\section{Additional details for Eqs.~(2) and (3)}
The detailed description of the charge self-consistency procedure is given in Ref.~\cite{shick09}.
In brief, once the occupation
matrix {$n_{\gamma_1 \gamma_2}$} is evaluated from the solution of Eq.~(1), it is used to
construct the effective ``LDA+U potential'',
${V}_{U} = \sum_{\gamma_1 \gamma_2}
  |\phi_{\gamma_1} \rangle V_{U}^{\gamma_1 \gamma_2} \langle
  \phi_{\gamma_2}|$,
where
\renewcommand{\theequation}{S1}
\begin{align}
 V_{U}^{\gamma_1 \gamma_2} =& \sum_{\gamma \gamma'}
 \Big(\langle \gamma_2 \gamma  |V^{ee}|\gamma_1  \gamma' \rangle
 - \langle \gamma_2 \gamma |V^{ee}| \gamma'  \gamma_1 \rangle
 \Big)
n_{\gamma \gamma'} \nonumber\\
&- V_{dc} \delta_{\gamma_1 \gamma_2}\,.
\label{eq:ldau}
\end{align}
Up to now, our considerations did not depend on the choice of the
basis set. The method becomes basis-set dependent when a projector of
the Bloch-state solutions of Eq.~(2) on
the local basis $\{ \phi_{\gamma} \}$ is specified. The FP-LAPW
method uses a basis set of plane waves that are matched onto a
linear combination of all radial solutions (and their energy
derivative) inside a sphere centered on each atom. In this case, we
make use of the projector technique which is described in detail in
Ref.~\cite{shick99}. It is important to mention that, due to the
full-potential character of the method, care should be taken to exclude
the double-counting of the $f$-states non-spherical contributions to
the LDA and LDA+U parts of potential in Eq.~(2).

The selfenergy $\Sigma$ entering  Eq.~(3) is obtained as
\renewcommand{\theequation}{S2}
\begin{equation}
\Sigma
=\hat G_{\rm imp}^{-1}\bigl[\hat H^{(0)}_{\rm imp}\bigr]
 - \hat G_{\rm imp}^{-1}\bigl[
\hat H_{\rm imp} 
\bigr]\,,
\end{equation}
where $\hat G_{\rm imp}[\hat H_{\rm imp}]$ represents the Green's function
matrix in the $f$-orbital subspace evaluated for a general impurity
Hamiltonian defined in Eq.~(1). The matrix $\hat G_{\rm imp}\bigl[\hat
H^{(0)}_{\rm imp}\bigr]$ is the Green's function of the
same Hamiltonian without the Coulomb interaction (fifth term).

\section{Magnetic susceptibility in the impurity model}

It is indicated in the main text that the magnetic susceptibility is a
decreasing function of the Hund's exchange $J$. Figure~\ref{fig:susc}
illustrates this dependence in detail for the crystal-field theory
($\chi_{\rm CF}$) as well as for the impurity model with hybridization
($\chi_{\rm imp}$). In principle it is possible to reduce the
susceptibility down to the experimental value if $J$ is raised to 1.0 eV
but such $J$ seems unrealistically high since it is about 20\% larger
than the atomic Hartree--Fock value \cite{KMoore2009}.

\renewcommand{\thefigure}{S3}
\begin{figure}[htbp]
\includegraphics[angle=0,width=0.975\columnwidth]{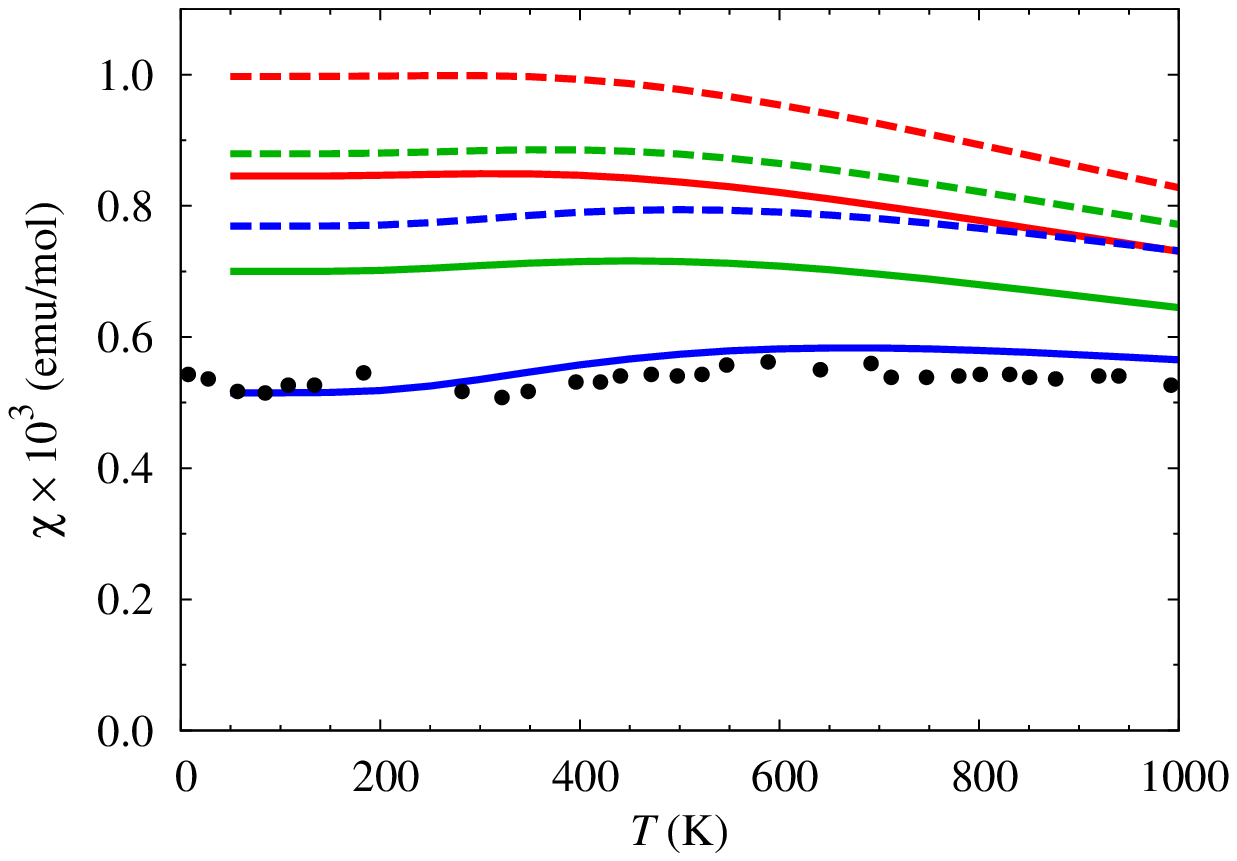}
\caption{(Color online) 
Magnetic susceptibility as a function of the Hund's exchange; $J=0.5$
eV (red), 0.65 eV (green) and 1.0 eV (blue). We compare $\chi_{\rm
  imp}$ from the impurity model (solid lines) with $\chi_{\rm CF}$
from the crystal-field theory (dashed lines). Experimental data
\cite{raphael1968} are shown as black dots.}
\label{fig:susc}
\end{figure}

The suppressed temperature dependence is
a result of cancellation of temperature dependent parts of the induced
moments $\mu_{\rm B}\langle\hat l_z\rangle$ and
$\mu_{\rm B}\langle 2\hat s_z\rangle$ that both deviate from a constant
above 300 K due to increasing population of the excited-state
triplet $\Gamma_4$. Figure~\ref{fig:suscTdep} illustrates this
cancellation: the susceptibility is split into two parts, $\chi_{\rm
  imp}=\chi_l+\chi_s$, where $\chi_l$ measures the response of the orbital
moment, $\chi_l(T)=\mu_0 N_{\rm A} \mu_{\rm B}\langle\hat
l_z\rangle/B_z$, and $\chi_s$ measures the response of the spin
moment, $\chi_s(T)=\mu_0 N_{\rm A} \mu_{\rm B}\langle 2\hat
s_z\rangle/B_z$.

The effect of hybridization with ligand states can be approximately
represented in the crystal-field theory by means of the so-called
orbital reduction factor $k$ that enters the Zeeman term in the
hamiltonian
\begin{equation*}
{\hat H}^f_Z =
 - \sum_{m\sigma}\mu_{\rm B} B_z
\bigl[k \hat l_z+2\hat s_z\bigr]_{mm}^{\sigma\;\;\sigma}
\hat f_{m\sigma}^{\dagger}\hat f_{m\sigma}
\end{equation*}
and also the magnetic moment induced in the $f$ shell $m_z=\mu_{\rm B}\langle k \hat
l_z+2\hat s_z\rangle$. The susceptibilities $\chi_{\rm imp}$ shown in
Fig.~\ref{fig:susc} can be recovered in the crystal-field theory by
setting $k$ to 0.965 for $J=0.5$ eV, 0.955 for $J=0.65$ eV, and 0.932 for
$J=1.0$ eV. The experimental data would correspond to $k=0.905$
for $J=0.65$ eV as shown in \cite{kern1999}.

\renewcommand{\thefigure}{S4}
\begin{figure}[htbp]
\includegraphics[angle=0,width=0.975\columnwidth]{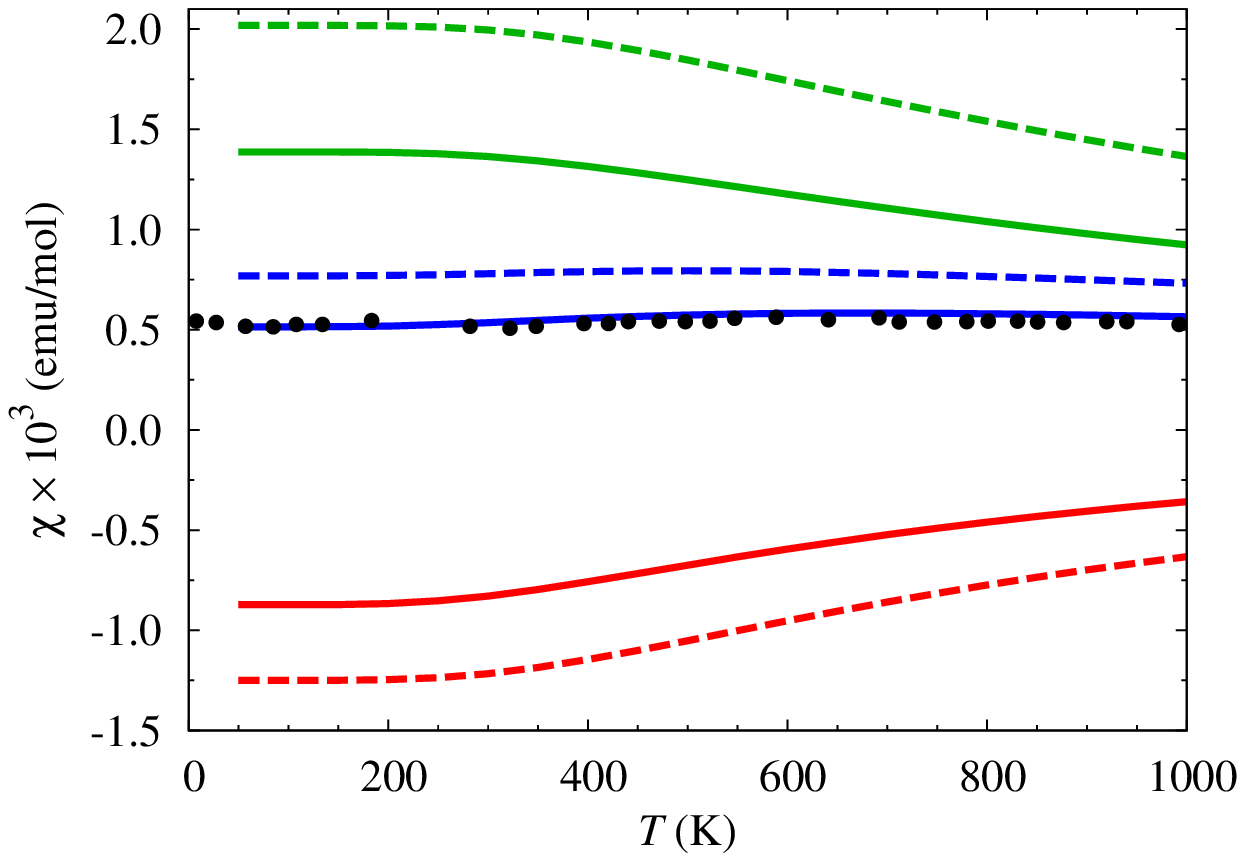}
\caption{(Color online)
Cancellation of the temperature dependence in the total magnetic
susceptibility (blue) for $J=1.0$ eV. The orbital
part $\chi_l$ is shown in green, the spin part $\chi_s$ in
red. Solid lines correspond to $\chi_{\rm imp}$, dashed lines to
$\chi_{\rm CF}$.}
\label{fig:suscTdep}
\end{figure}

\end{document}